\documentclass{article}%
\usepackage{amsmath}
\usepackage{amsfonts}
\usepackage{amssymb}
\usepackage{graphicx}%
\begin{document}

\title{Constructive use of holographic projections\\{\small Dedicated to Klaus Fredenhagen on the occasion of his 60th}
{\small birthday}}
\author{Bert Schroer\\CBPF, Rua Dr. Xavier Sigaud 150 \\22290-180 Rio de Janeiro, Brazil\\and Institut fuer Theoretische Physik der FU Berlin, Germany}
\date{November 2007}
\maketitle

\begin{abstract}
Revisiting the old problem of existence of interacting models of QFT with new
conceptual ideas and mathematical tools, one arrives at a novel view about the
nature of QFT. The recent success of algebraic methods in establishing the
existence of factorizing models suggests new directions for a more intrinsic
constructive approach beyond Lagrangian quantization. Holographic projection
simplifies certain properties of the bulk theory and hence is a promising new
tool for these new attempts.

To appear in: "Quantum field theory --Competitive Methods"

Birkhh\"{a}user Basel, 2008, Eds. B. Fauser, J. Tolksdorf and E. Zeidler

\end{abstract}

\section{Historical background and present motivations for holography}

No other theory in the history of physics has been able to cover such a wide
range of phenomena with impressive precision as QFT. However its amazing
predictive power stands in a worrisome contrast to its weak ontological
status. In fact QFT is the only theory of immense epistemic strength which,
even after more than 80 years, remained on shaky mathematical and conceptual
grounds. Unlike any other area of physics, including QM, there are simply no
interesting mathematically controllable interacting models which would show
that the underlying principles remain free of internal contradictions in the
presence of interactions. The faith in e.g. the Standard Model is based
primarily on its perturbative descriptive power; outside the perturbative
domain there are more doubts than supporting arguments.

The suspicion that this state of affairs may be related to the conceptual and
mathematical weakness of the method of Lagrangian quantization rather then a
shortcoming indicating an inconsistency of the underlying principles in the
presence of interactions can be traced back to its discoverer Pascual Jordan.
It certainly was behind all later attempts of e.g. Arthur Wightman and Rudolf
Haag to find a more autonomous setting away from the quantization parallelism
with classical theories which culminated in Wightman's axiomatic setting in
terms of vacuum correlation functions and the Haag-Kastler theory of nets of
operator algebras.\ 

The distance of such conceptual improvements to the applied world of
calculations has unfortunately persisted. Nowhere is the contrast between
computational triumph and conceptual misery more visible than in renormalized
perturbation theory which has remained our only means to explore the standard
model. Most particle physicists have a working knowledge of perturbation
theory and at least some of them took notice of the fact that, although the
renormalized perturbative series can be shown to diverge and that in certain
cases these divergent series are Borel resummable. Here I will add some more
comments without going into details.

The Borel re-summability property unfortunately does not lead to an existence
proof; the correct mathematical statement in this situation is that if the
existence can be established\footnote{The existence for models with a finite
wave-function renormalization constant has been established in the early 60s
and this situation has not changed up to recently. The old results only
include superrenormalizable models whereas the new criterion is not related to
short-distance restrictions but rather requires a certain phase space behavior
(modular nuclearity).} by nonperturbative method then the Borel-resummed
series would indeed acquire an asymptotic convergence status with respect to
the solution, and one would for the first time be allowed to celebrate the
numerical success as having a solid ontological basis \footnote{This is
actually the present situation for the class of d=1+1 factorizing models
\cite{Lech}.}. But the whole issue of model existence attained the status of
an unpleasant fact, something which is often kept away from newcomers, so that
as a result there is a certain danger to confuse the existence of a model with
the ability to write down a Lagrangian or a functional integral and apply some
computational recipe.

Fortunately important but unfashionable problems in particle physics never
disappear completely. Even if they have been left on the wayside as
"un-stringy", "unsupersymmetrizable" or too far removed from the "Holy Grail
of a TOE" and therefore not really career-improving, there will be always be
individuals who return to them with new ideas.

Indeed there has been some recent progress about the aforementioned existence
problem from a quite unexpected direction. Within the setting of d=1+1
factorizing models the use of modular operator theory has led to a control
over phase space degrees of freedom which in turn paved the way to an
existence proof. Those models are distinguished by their simple generators for
the wedge-localized algebra \cite{Sch}; in fact these generators turned out to
possess Fourier-transforms with mass-shell creation/annihilation operators
which are only slightly more complicated than free fields. An important
additional idea on the way to an existence proof is the issue of the
cardinality of degrees of freedom. In the form of the phase space in QFT as
opposed to QM this issue goes back to the 60s \cite{Ha-Sw} and underwent
several refinements \cite{Bu-Wi} (a sketch of the history can be found in
\cite{legacy}).

The remaining problem was to show that the simplicity of the wedge generators
led to a "tame" phase space behavior which guaranties the nontriviality as
well as the additional expected properties of the double cone localized
algebras obtained as intersections of wedge-localized algebras \cite{Lech}.
Although these models have no particle creation through on-shell scattering,
they exhibit the full infinite vacuum polarization clouds upon sharpening the
localization from wedges to compact spacetime regions as e.g. double cones
\cite{Bo-Bu-Sc}. Their simplicity is only manifest in the existence of simple
wedge generators; for compact localization regions their complicated infinite
vacuum polarization clouds are not simpler than in other QFT. 

Similar simple-minded Ans\"{a}tze for wedge algebras in higher dimensions
cannot work since interactions which lead to nontrivial elastic scattering
without also causing particle creation cannot exist; such a No-Go theorem for
4-dim. QFT was established already in \cite{Aks}. Nevertheless it is quite
interesting to note that even if with such a simple-minded Ansatz for wedge
generators in higher dimensions one does not get to compactly localized local
observables, one can in some cases go to certain subwedge intersections
\cite{Bu-Su}\cite{Gr-Le} before the increase in localization leads to trivial algebras.

Whereas in the Lagrangian approach one starts with local fields and their
correlations and moves afterwards to less local objects as global charges,
incoming fields\footnote{Incoming/outgoing free fields are only local with
respect to themselves. The physically relevant notion of locality is
\textit{relative locality to the interacting fields}. If incoming fields are
relatively local/almost local, the theory has no interactions. } etc., the
modular localization approach goes the opposite way i.e. one starts from the
wedge region (the best compromise between particles and fields) which is most
close to the particle mass-shell the S-matrix and then works one's way down.
The pointlike local fields only appear at the very end and play the role of
\textit{coordinatizing generators} of the double cone algebras for arbitrary
small sizes. 

Nonlocal models are automatically "noncommutative" in the sense that the the
maximal commutativity of massive theories allowed by the principles of QFT,
namely spacelike commutativity, is weakened by allowing various degrees of
violations of spacelike commutativity. In this context the non-commutativity
associated with the deformation of the product to a star-product using the
Weyl-Moyal formalism is only a very special (but very popular) case. The
motivation for studying non-commutative QFT for its own sake comes from string
theory, and one should not expect this motivation to be better than for string
theory itself. 

My motivation for having being interested in noncommutative theory during the
last decade comes from the observation that non-commutative fields can have
\textit{simpler properties than commutative ones}. More concretely:
complicated two-dimensional local theories may lead to wedge-localized
algebras which are generated by non-commutative fields where the latter only
fulfill the much weaker wedge-locality (see above). Whereas in d=1+1 such
constructions \cite{Sch} may lead via algebraic intersections to nontrivial,
nonperturbative local fields, it is known that in higher dimensions this
simple kind of wedge generating field without vacuum polarization is not
available. But interestingly enough on can improve the wedge localization
somewhat \cite{new} before the further sharpening of localization via
algebraic intersections ends in trivial algebras.

These recent developments combine the useful part of the history of S-matrix
theory and formfactors with very new conceptual inroads into QFT (modular
localization, phase space properties of LQP). \ The idea to divide the
difficult full problem into a collection of simpler smaller ones is also at
the root of the various forms of the holography of the two subsequent sections.

The predecessor of lightfront holography was the so-called "lightcone
quantization" which started in the early 70s; it was designed to focus on
short-distances and forget temporarily about the rest. The idea to work with
fields which are associated to the lightfront $x_{-}=0$ (not the light cone
which is $x^{2}=0$) as a submanifold in Minkowski spacetime looked very
promising but unfortunately the connection with the original problem of
analyzing the local theory in the bulk was never addressed and as the
misleading name "lightcone quantization" reveals, the approach was considered
as a different quantization rather then a different method for looking at the
same local QFT in Minkowski spacetime. It is not really necessary to continue
a seperate criticism of "lightcone quantization" because its shortcomings will
be become obvious after the presentation of lightfront holography (more
generally \textit{holography onto null-surfaces}).

Whereas the more elaborate and potentially more important lightfront
holography has not led to heated discussions, the controversial potential of
the simpler AdS-CFT holography had been enormous and to the degree that it
contains interesting messages which increase our scientific understanding it
will be presented in these notes.

Since all subjects have been treated in the existing literature, our
presentation should be viewed ass a guide through the literature with
occasionally additional and (hopefully) helpful remarks. 

\section{Lightfront holography, holography on null-surfaces and the origin of
the area law}

Free fields offer a nice introduction into the bulk-holography relation which,
despite its simplicity, remains conceptually non-trivial.

We seek generating fields $A_{LF}$ for the lightfront algebra $\mathcal{A}%
(LF)$ by following the formal prescription $x_{-}=0$ of the old "lightfront
approach" \cite{Leut}. Using the abbreviation $x_{\pm}=x^{0}\pm x^{3},~p_{\pm
}=p^{0}+p^{3}\simeq e^{\mp\theta},~$with $\theta$ the momentum space rapidity :%

\begin{align}
&  A_{LF}(x_{+},x_{\perp}):=A(x)|_{x_{-}=0}\simeq\int\left(  e^{i(p_{-}%
(\theta)x_{+}+ip_{\perp}x_{\perp}}a^{\ast}(\theta,p_{\perp})d\theta dp_{\perp
}+h.c.\right) \label{LF}\\
&  \left\langle \partial_{x_{+}}A_{LF}(x_{+},x_{\perp})\partial_{x\prime_{+}%
}A_{LF}(x_{+}^{\prime},x_{\perp}^{\prime})\right\rangle \simeq\frac{1}{\left(
x_{+}-x_{+}^{\prime}+i\varepsilon\right)  ^{2}}\cdot\delta(x_{\perp}-x_{\perp
}^{\prime})\nonumber\\
&  \left[  \partial_{x_{+}}A_{LF}(x_{+},x_{\perp}),\partial_{x\prime_{+}%
}A_{LF}(x_{+}^{\prime},x_{\perp}^{\prime})\right]  \simeq\delta^{\prime}%
(x_{+}-x_{+}^{\prime})\delta(x_{\perp}-x_{\perp}^{\prime})\nonumber
\end{align}
The justification for this formal manipulation\footnote{We took the
derivatives for technical reasons (in order to write the formulas without test
functions).} follow from the fact that the equivalence class of test function
$\left[  f\right]  $, which have the same mass shell restriction $\tilde
{f}|_{H_{m}}$ to the mass hyperboloid of mass m, is mapped to a unique test
function $f_{LF}$ which "lives"on the lightfront \cite{Dries}\cite{S1}. It
only takes the margin of a newspaper to verify the identity $A(f)=A(\left[
f\right]  )=A_{LF}(f_{LF}).$ This identity does not mean that the
\thinspace$A_{LF}$ generator can be used to describe the local substructure in
the bulk. The inversion involves an equivalence class and does not distinguish
an individual test-function in the bulk; in fact a finitely localized test
function $f(x_{+},x_{\perp})$ on LF corresponds to a de-localized subspace in
the bulk. Using an intuitive metaphoric language one may say that a strict
localization on LF corresponds to a fuzzy localization in the bulk and vice
versa. Hence the pointwise use of the LF generators enforces the LF
localization and the only wedge-localized operators which can be directly
obtained as smeared $A_{LF}$ fields have a noncompact extension within a wedge
whose causal horizon is on LF. Nevertheless there is equality between the two
operator algebras associated to the bulk $W$ and its (upper) horizon $\partial
W$%
\begin{equation}
\mathcal{A}(W)=\mathcal{A}(H(W))\subset\mathcal{A}(LF)=B(H)
\end{equation}
These operator algebras are the von Neumann closures of the Weyl algebras
generated by the smeared fields $A$ and $A_{LF}$ and it is only in the sense
of this closure (or by forming the double commutant) that the equality holds.
Quantum field theorists are used to deal with single operators. Therefore the
knowledge about the equality of algebras without being able to say which
operators are localized in subregion is somewhat unaccustomed. As will be
explained later on, the finer localization properties in the algebraic setting
can be recovered by taking suitable intersections of wedge algebras i.e. the
structure of the family of all wedge algebras determines whether the local
algebras are nontrivial and in case they are permits to compute the local net
which contains all informations about the particular model.

This idea of taking the holographic projection of individual bulk fields can
be generalized to composites of free fields (as e.g. the stress-energy
tensor). In order to avoid lengthy discussions about how to interpret
logarithmic chiral two-point functions in terms of restricted test
functions\footnote{This is a well-understood problem of chiral fields of zero
scale dimension which is not directly related to holography.} we work restrict
our attention to Wick-composites of $\partial_{x_{+}}A_{LF}(x_{+},x_{\perp})$
\begin{equation}
\left[  B_{LF}(x_{+},x_{\perp}),C_{LF}(x_{+}^{\prime},x_{\perp}^{\prime
})\right]  =\sum_{l=0}^{m}\delta^{l}(x_{\perp}-x_{\perp}^{\prime}%
)\sum_{k(l)=0}^{n(l)}\delta^{k(l)}(x_{+}-x_{+}^{\prime})D_{LF}^{\left(
k(l)\right)  }(x_{+},x_{\perp})\label{com}%
\end{equation}
where the dimensions of the composites $D_{LF}^{(k(l))}$ together with the
degrees of the derivatives of the delta functions obey the standard rule of
scale dimensional conservation. In the commutator the transverse and the
longitudinal part both appear with delta functions and their derivatives yet
there is a very important structural difference which shows up in the
correlation functions. To understand this point we look at the second line in
(\ref{LF}). The longitudinal (=lightlike) delta-functions carries the chiral
vacuum polarization the transverse part consists only of products of delta
functions as if it would come from a product of correlation functions of
nonrelativistic Schroedinger creation/annihilation operators $\psi^{\ast
}(x_{\perp}),$ $\psi(x_{\perp}).$ In other words the LF-fields which feature
in this extended chiral theory are \textit{chimera between QFT and QM}; they
have one leg in QFT and n-2 legs in QM with the "chimeric vacuum" being
partially a (transverse) factorizing quantum mechanical state of "nothingness"
(the Buddhist nirvana) and partially the longitudinally particle-antiparticle
polarized LQP vacuum state of "virtually everything" (the Abrahamic heaven).

Upon lightlike localization of LF to (in the present case) $\partial W$ (or to
a longitudinal interval) the vacuum on $\mathcal{A}(\partial W)$ becomes a
radiating KMS thermal state with nonvanishing localization-entropy
\cite{S1}\cite{S2}. In case of interacting fields there is no change with
respect to the absence of transverse vacuum polarization, but unlike the free
case the global algebra $\mathcal{A}(LF)$ or the semi-global algebra
$\mathcal{A}(\partial W)$ is generally bigger than the algebra one obtains
from the globalization using compactly localized subalgebras, i.e.
$\overline{\cup_{O\subset LF}\mathcal{A}_{LF}(\mathcal{O})}\subset
\mathcal{A}(LF),$ $\mathcal{O}\subset LF$. We will return to this point at a
more opportune moment.

The aforementioned "chimeric" behavior of the vacuum is related in a profound
way to the conceptual distinctions between QM and QFT \cite{interface}.
Whereas transversely the vacuum is tensor-factorizing with respect to the Born
localization and therefore leads to the standard quantum mechanical concepts
of entanglement and the related information theoretical (cold) entropy, the
entanglement from restricting the vacuum to an algebra associated with an
interval in lightray direction is a thermal KMS state with a genuine
thermodynamic entropy. Instead of the standard quantum mechanical dichotomy
between pure and entangled restricted states there are simply no pure states
at all. All states on sharply localized operator algebras are highly mixed and
the restriction of global particle states (including the vacuum) to the
W-horizon $\mathcal{A}(\partial W)$ results in KMS thermal states. This is the
result of the different nature of localized algebras in QFT from localized
algebras in QM \cite{interface}.

Therefore if one wants to use the terminology\ "entanglement" in QFT one
should be aware that one is dealing with a totally intrinsic very strong form
of entanglement: \textit{all physically distinguished global pure states} (in
particular finite energy states in particular the vacuum) \textit{upon
restriction to a localized algebra become intrinsically entangled and unlike
in QM there is no local operation which disentangles}.

Whereas the cold (information theoretic) entanglement is often linked to the
uncertainty relation of QM, the raison d'etre behind the "hot" entanglement is
the phenomenon of vacuum polarization resulting from localization in quantum
theories with a maximal velocity. The transverse tensor factorization
restricts the Reeh-Schlieder theorem (also known as the "state-operator
relation"). For a longitudinal strip (st) on LF of a finite transverse
extension the LF algebra tensor-factorizes together with the Hilbert space
$H=H_{st}\otimes H_{st\perp}$ and the $H_{st}$ projected form of the
Reeh-Schlieder theorem for a subalgebra localized within the strip continues
to be valid$.$

This concept of \textit{transverse extended chiral fields} can also be
axiomatically formulated for interacting fields independently of whether those
objects result from a bulk theory via holographic projection or whether one
wants to study QFT on (non-hyperbolic) null-surfaces. These "lightfront
fields" share some important properties with chiral fields. In both cases
subalgebras localized on subregions lead to a \textit{geometric modular
theory,} whereas in the bulk this property is restricted to wedge algebras.
Furthermore in both cases the symmetry groups are infinite dimensional; in
chiral theories the largest possible group is (after compactification)
Diff($\dot{R}$), whereas the transverse extended version admits besides these
pure lighlike symmetries also $x_{\perp}$-$x_{+}$ mixing ($x_{\perp}%
$-dependent) symmetry transformations which leave the commutation structure invariant.

There is one note of caution, unlike those conformal QFTs which arise as
chiral projections from 2-dimensional conformal QFT, the extended chiral
models of QFT on the lightfront which result from holography do not come with
a stress-energy tensor and hence the diffeomorphism invariance beyond the
Moebius invariance (for which one gets from modular invariance, no energy
momentum tensor needed) is not automatic. \ This leads to the interesting
question if there are concepts which permit to incorporate also the
diffeomorphisms beyond the Moebius transformations into a modular setting, a
problem which will not be pursuit here.

We have formulated the algebraic structure of holographic projected fields for
bosonic fields, but it should be obvious to the reader that a generalization
to Fermi fields is straightforward. Lightfront holography is consistent with
the fact that except for d=1+1 there are no operators which "live" on a
lightray since the presence of the quantum mechanical transverse delta
function prevents such a possibility i.e. only after transverse averaging with
test functions does one get to (unbounded) operators.

It is an interesting question whether a direct "holographic projection" of
\textit{interacting} pointlike bulk fields into lightfront fields analog to
(\ref{LF}) can be formulated, thus avoiding the algebraic steps starting with
wedge algebra. The important formula which led to the lightfront generators is
the \textit{mass shell representation} of the free field; if we would have
performed the $x_{-}=0$ limit in the two point function the result would
diverge. This suggests that we should start from the so-called
Glaser-Lehmann-Zimmermann (GLZ) representation \cite{GLZ} which is an on-shell
representation in terms an infinite series of integrals involving the incoming
particle creation/annihilation operators
\begin{align}
&  A(x)=\sum\frac{1}{n!}\int dx_{1}...\int dx_{n}~a(x;x_{1},...x_{n}%
):A_{in}(x_{1})....A(x_{n}):\label{mass}\\
&  A(x)=\sum\frac{1}{n!}\int_{H_{m}}dp_{1}...\int_{H_{m}}dp_{n}~e^{ix(\sum
p_{i})}\tilde{a}(p_{1},...p_{n}):\tilde{A}(p_{1})....\tilde{A}(p_{n}%
):\nonumber\\
&  A(x)_{LF}=A(x)_{x_{-}=0}\nonumber
\end{align}
in which the coefficient functions $a(x;x_{1},...x_{n})$ are \textit{retarded
functions}. The second line shows that only the mass-shell restriction of
these functions matter; the momentum space integration goes over the entire
mass-shell and the two components of the mass hyperboloid $H_{m}$ are
associated with the annihilation/creation part of the Fourier transform of the
incoming field. These mass-shell restrictions of the retarded coefficient
functions are related to multi-particle formfactors of the field $A.$ Clearly
we can take $x_{-}=0$ in this on-shell representation without apparently
creating any problems in addition to the possibly bad convergence properties
of such series (with or without the lightfront restriction) which they had
from the start. The use of the on-shell representation (\ref{mass}) is
essential, doing this directly in the Wightman functions would lead to
meaningless divergences, as we already noticed in the free field case.

Such GLZ formulas amount to a representation of a local field in terms of
other local fields in which the \textit{relation between the two sets} of
fields is very \textit{nonlocal}. Hence this procedure is less intuitive than
the algebraic method based on relative commutants and intersections of
algebras. The use of a GLZ series also goes in some sense against the spirit
of holography which is to simplify certain aspects\footnote{Those aspects for
which holography does not simplify include particle and scattering aspects.}
in order to facilitate the solution of certain properties of the theory (i.e.
to preserve the original aim of the ill-defined lightcone quantization),
whereas to arrive at GLZ representations one must already have solved the
on-shell aspects of the model (i.e. know all its formfactors) before applying holography.

Nevertheless, in those cases where one has explicit knowledge of formfactors,
as in the case of 2-dim. factorizing models mentioned in the previous section,
this knowledge can be used to calculate the scaling dimensions of their
associated holographic fields $A_{LF}.$ These fields lead to more general
plektonic (braid group) commutation relations which replace the bosonic
relations of transverse extended chiral observables (\ref{com}). We refer to
\cite{Hol} in which the holographic scaling dimensions for several fields in
factorizing models will be calculated, including the Ising model for which an
exact determination of the scaling dimension of the order field is possible.
Although the holographic dimensions agree with those from the short distance
analysis (which have been previously calculated in \cite{Ba-Ka}), the
conceptual status of holography is quite different from that of critical
universality classes. The former is an exact relation between a 2-dim.
factorizing model (change of the spacetime ordering of a given bulk theory)
whereas the latter is a passing to a different QFT in the same universality
class. The mentioned exact result in the case of the Ising model strengthens
the hope that GLZ representations and the closely related expansions of local
fields in terms of wedge algebra generating on-shell operators \cite{Hol} have
a better convergence status than perturbative series.

By far the conceptually and mathematically cleanest way to pass from the bulk
to the lightfront is in terms of nets of operator algebras via modular theory.
This method requires to start from algebras in "standard position" i.e. a pair
($\mathcal{A},\Omega$) such that the operator algebra $\mathcal{A}$ acts
cyclically on the state vector $\Omega$ i.e. $\overline{\mathcal{A}\Omega}=H$
and has no annihilators i.e. $A\Omega=0\curvearrowright A=0.$ According to the
Reeh-Schlieder theorem any localized algebra $\mathcal{A}(\mathcal{O})$ forms
a standard pair ($\mathcal{A}(\mathcal{O}),\Omega$) with respect to the vacuum
$\Omega$ and the best starting point for the lightfront holography is a wedge
algebra since the (upper) causal horizon $\partial W$ of the wedge $W$ is
already half the lightfront. The crux of the matter is the construction of the
local substructure on $\partial W.$ The local resolution in longitudinal
(lightray) direction is done as follows.

Let $W$ be the $x_{0}-x_{3}$ wedge in Minkowski spacetime which is left
invariant by the $x_{0}-x_{3}$ Lorentz-boosts. Consider a family of wedges
$W_{a}$ which are obtained by sliding the $W$ along the $x_{+}=x_{0}+x_{3}$
lightray by a lightlike translation $a>0$ into itself. The set of spacetime
points on $LF$ consisting of those points on $\partial W_{a}$ which are
spacelike to the interior of $W_{b}$ for $b>a$ is denoted by $\partial
W_{a,b};$ it consists of points $x_{+}\in(a,b)$ with an unlimited transverse
part $x_{\perp}\in R^{2}$. These regions are two-sided transverse slabs on
$LF$.

To get to intersections of finite size one may \textquotedblleft
tilt\textquotedblright\ these slabs by the action of certain subgroups in
$\mathcal{G}$ which change the transverse directions. Using the 2-parametric
subgroup $\mathcal{G}_{2}$ of $\mathcal{G}$ which is the restriction to $LF$
of the two \textquotedblleft translations\textquotedblright\ in the Wigner
little group (i.e. the subgroup fixing the lightray in $LF$), it is easy to
see that this is achieved by forming intersections with $G_{2}$- transformed
slabs $\partial W_{a,b}$
\begin{equation}
\partial W_{a,b}\cap g(\partial W_{a,b}),\text{ }g\in\mathcal{G}_{2}%
\end{equation}
By continuing with forming intersections and unions, one can get to finite
convex regions $\mathcal{O}$ of a quite general shape.

The local net on the lightfront is the collection of all local algebras
$\mathcal{A(O}),$ $\mathcal{O}\subset LF$ and as usual their weak closure is
the global algebra $\mathcal{A}_{LF}$. For interacting systems the global
lightfront algebra is generally expected to be smaller than the bulk, in
particular one expects
\begin{align}
\mathcal{A}_{LF}(\partial W) &  \subset\mathcal{A}(\partial W)=\mathcal{A}%
(W)\label{contain}\\
\mathcal{A}_{LF}(\partial W) &  =\cup_{\mathcal{O}\subset\partial
W}\mathcal{A}_{LF}(\mathcal{O}),~\mathcal{A}(W)=\cup_{\mathcal{C\subset}%
W}\mathcal{A}(\mathcal{C})\nonumber
\end{align}
where the semi-global algebras are formed with the localization concept of
their relative nets as indicated in the second line. The smaller left hand
side accounts for the fact that the formation of relative commutants as
$\mathcal{A}(\partial W_{a,b})$ may not maintain the standardness of the
algebra because $\overline{\cup_{a,b}\mathcal{A}(\partial W_{a,b})\Omega
}\subsetneqq H.$ In that case the globalization of the algebraic holography
only captures a global (i.e. not localized) subalgebra of the global bulk and
one could ask whether the pointlike procedure using the GLZ representation
leads to generating fields which generate a bigger algebra. gives more. The
answer is positive since also (bosonic) fields with anomalous short distance
dimensions will pass the projective holography and become anyonic field on the
lightray\footnote{The standard Boson-Fermion statistics refers to spacelike
distances and the lightlike statistics resulting from projective holography is
determined by the anomalous short distance dimensions of the bulk field and
not by their statistics.}  On the other hand algebraic holography filters out
bosonic fields which define the chiral obervables. These chiral observables
have a DHR superselection theory. This leads to the obvious conjecture%
\begin{equation}
Alg\{proj~hol\}\subseteq Alg\{DHR\}
\end{equation}
Here the left hand side denotes the algebra generated by applying projective
holography to the pointlike bulk fields and the reight hand side is the
smallest algebra which contains all DHR superselection sectors of the LF
observable (extended chiral) algebra which resulted from algebraic holography.

It is worthwhile to emphasize that the connection between the operator
algebraic and the pointlike prescription is much easier on LF than in the
bulk. In the presence of conformal symmetries one has the results of Joerss
\cite{Joerss}; looking at his theorems in the chiral setting an adaptation to
the transverse extended chiral theories on LF should be straightforward. For
consistency reasons such fields must fulfill (\ref{com}) I hope to come back
to this issue in a different context.

One motivation for being interested in lightfront holography is that it is
expected to helpful in dividing the complicated problem of classifying and
constructing QFTs according to intrinsic principles into several less
complicated steps. In the case of d=1+1 factorizing models one does not need
this holographic projection onto a chiral theory on the lightray for the mere
existence proof. But e.g. for the determination of the spectrum of the short
distance scale dimension, it is only holography and not the critical limit
which permits to maintain the original Hilbert space setting. It is precisely
this property which makes it potentially interesting for structural
investigations and actual constructions of higher dimensional QFT.

Now we are well-prepared to address the main point of this section: the area
law for localization entropy which follows from the absence of transverse
vacuum polarization. Since this point does not depend on most of the above
technicalities, it may be helpful to the reader to present the conceptual
mathematical origin of this unique\footnote{Holography on null-surfaces is the
only context in which a quantum mechanical structure enters a field theoretic
setting.} tensor-factorization property. The relevant theorem goes back to
Borchers \cite{Bo} and can be stated as follows. Let $\mathcal{A}_{i}\subset
B(H),$ $i=1,2$ be two operator algebras with $\left[  \mathcal{A}%
_{1},U(a)\mathcal{A}_{2}U(a)^{\ast}\right]  =0$ $\forall a$ and $U(a)$ a
translation with \textit{nonnegative} generator which fulfills the cluster
factorization property (i.e. asymptotic factorization in correlation functions
for infinitely large cluster separations) with respect to a unique
$U(a)$-invariant state vector $\Omega$\footnote{Locality in both directions
shows that the lightlike translates $\left\langle \Omega\left\vert
AU(a)B\right\vert \Omega\right\rangle $ are boundary values of entire
functions and the cluster property together with Liouville's theorem gives the
factorization.}$.$ It then follows that the two algebras tensor factorize in
the sense $\mathcal{A}_{1}\mathcal{\vee A}_{2}=\mathcal{A}_{1}\mathcal{\otimes
A}_{2}$ where the left hand side denotes the joint operator algebra.

In the case at hand the tensor factorization follows as soon as the open
regions $\mathcal{O}_{i}\subset LF$ in $\mathcal{A(O}_{i}$) $i=1,2$ have no
transverse overlap. The lightlike cluster factorization is weaker (only a
power law) than its better known spacelike counterpart, but as a result of the
analytic properties following from the \textit{non-negative generator of
lightlike translations} it enforces the asymptotic factorization to be valid
at all distances. The resulting transverse factorization implies the
transverse additivity of extensive quantities as energy and entropy and their
behavior in lightray direction can then be calculated in terns of the
associated auxiliary chiral theory. a well-known property for spacelike separations.

This result \cite{S1}\cite{S2} of the transverse factorization may be
summarized as follows

\begin{enumerate}
\item The system of $LF$ subalgebras $\left\{  \mathcal{A(O)}\right\}
_{\mathcal{O\subset}LF}$ tensor-factorizes transversely with the vacuum being
free of transverse entanglement
\begin{align}
&  \mathcal{A(O}_{1}\mathcal{\cup O}_{2}\mathcal{)}=\mathcal{A(O}%
_{1}\mathcal{)\otimes A(O}_{2}\mathcal{)},\text{ }\mathcal{(O}_{1}%
\mathcal{)}_{\perp}\cap\mathcal{(O}_{2}\mathcal{)}_{\perp}=\emptyset
\label{fac}\\
&  \left\langle \Omega\left\vert \mathcal{A(O}_{1}\mathcal{)\otimes A(O}%
_{2}\mathcal{)}\right\vert \Omega\right\rangle =\left\langle \Omega\left\vert
\mathcal{A(O}_{1}\mathcal{)}\left\vert \Omega\right\rangle \left\langle
\Omega\right\vert \mathcal{A(O}_{2}\mathcal{)}\right\vert \Omega\right\rangle
\nonumber
\end{align}

\item Extensive properties as entropy and energy on $LF$ are proportional to
the extension of the transverse area.

\item The area density of localization-entropy in the vacuum state for a
system with sharp localization on $LF$ diverges logarithmically
\begin{equation}
s_{loc}=\lim_{\varepsilon\rightarrow0}\frac{c}{6}\left\vert ln\varepsilon
\right\vert +... \label{ent}%
\end{equation}
where $\varepsilon$ is the size of the interval of \textquotedblleft
fuzziness\textquotedblright\ of the boundary in the lightray direction which
one has to allow in order for the vacuum polarization cloud to attenuate and
the proportionality constant $c$ is (at least in typical examples) the central
extension parameter of the Witt-Virasoro algebra.
\end{enumerate}

The following comments about these results are helpful in order to appreciate
some of the physical consequences as well as extensions to more general null-surfaces.

As the volume divergence of the energy/entropy in a heat bath thermal system
results from the thermodynamic limit of a sequence of boxed systems in a Gibbs
states, the logarithmic divergence in the vacuum polarization attenuation
distance $\varepsilon$ plays an analogous role in the approximation of the
semiinfinitely extended $\partial W$ by sequences of algebras whose
localization regions approach $\partial W$ from the inside. In both cases the
limiting algebras are monads whereas the approximands are type I \ analogs of
the "box quantization" algebras. In fact in the present conformal context the
relation between the standard heat bath thermodynamic limit and the limit of
vanishing attenuation length for the localization-caused vacuum polarization
cloud really gord beyond an analogy and becomes an isomorphism.

This surprising result is based on two facts \cite{S1}\cite{S2}. On the one
hand conformal theories come with a natural covariant "box" approximation of
the thermodynamic limit since the continuous spectrum translational
Hamiltonian can be obtained as a scaled limit of a sequence of discrete
spectrum conformal rotational Hamiltonians associated to global type I
systems. In the other hand it has been known for some time that a heat bath
chiral KMS state can always be re-interpreted as the Unruh restriction applied
to a vacuum system in an larger world (a kind of inverse Unruh effect). Both
fact together lead to the above formula for the area density of entropy. In
fact using the conformal invariance one can write the area density formula in
the more suggestive manner by identifying $\varepsilon$ with the conformal
invariant cross-ratio of 4 points%
\[
\varepsilon^{2}=\frac{\left(  a_{2}-a_{1}\right)  \left(  b_{1}-b_{2}\right)
}{\left(  b_{1}-a_{1}\right)  \left(  b_{2}-a_{2}\right)  }%
\]
where $a_{1}<a_{2}<b_{2}<b_{1}$ so that $\left(  a_{1},b_{1}\right)  $
corresponds to the larger localization interval and $\left(  a_{2}%
,b_{2}\right)  $ is the approximand which goes with the interpolating type I
algebras. At this point one makes contact with some interesting work on what
condensed matter physicist call the "entanglement entropy"\footnote{In
\cite{Cardy} the formula for the logarithmically increasing entropy is
associated with a field theoretic cutoff and the role of the vacuum
polarization cloud in cunjunction with the KMS thermal properties (which is
not compatible with a quantum mechanical entanglement interpretation
\cite{interface}) are not noticed. Since there is no implementation of the
split property, the idea of an attenuation of the vacuum polarization cloud
has no conceptual place in a path integral formulation. QM and QFT are not
distinguished in the functional integral setting and even on a metaphorical
level there seems to be no possibility to implement the split property. }. 

One expects that the arguments for the absence of transverse vacuum
fluctuations carry over to other null-surfaces as e.g. the upper horizon
$\partial\mathcal{D}$ of the double cone $\mathcal{D}$. In the interacting
case it is not possible to obtain $\partial\mathcal{D}$ generators through
test function restrictions. For zero mass free fields there is however the
possibility to conformally transform the wedge into the double cone and in
this way obtain the holographic generators as the conformally transformed
generators of $\mathcal{A}(\partial W).$ In order to show that the resulting
$\mathcal{A(\partial D})$ continue to play their role even when the bulk
generators cease to be conformal one would have to prove that certain
double-cone affiliated inclusions are modular inclusions. We hope to return to
this interesting problem.

We have presented the pointlike approach and the algebraic approach next to
each other, but apart from the free field we have not really connected them.
Although one must leave a detailed discussion of their relation to the future,
there are some obvious observations one can make. Since for chiral fields the
notion of short-distance dimension and rotational spin (the action of the
$L_{0}$ generator) are closely connected and since the algebraic process of
taking relative commutators is bosonic, the lightfront algebras are
necessarily bosonic. A field as the chiral order variable of the Ising model
with dimension $\frac{1}{16}$ does not appear in the algebraic holography but,
as mentioned above, it is the pointlike projection of the massive order
variable in the factorizing Ising model in the bulk. On the other hand an
integer dimensional fields as the stress-energy tensor, is common to both
formulations. This suggests that the anomalous dimensional fields which are
missing in the algebraic construction may be recovered via representation
theory of the transverse extended chiral observable algebra which arises as
the image of the algebraic holography.

Since the original purpose of holography similar to that of that of its
ill-fated lightcone quantization predecessor, is to achieve a simplified but
still rigorous description (for the lightcone quantization the main motivation
was a better description of certain "short distance aspects" of QFT), the
question arises if one can use holography as a tool in a more ambitious
program of classification and construction of QFTs. In this case one must be
able to make sense of \textit{inverse holography} i.e. confront the question
whether, knowing the local net on the lightfront. one can only obtain at least
part of the local substructure of the bulk. It is immediately clear that one
construct that part in the bulk which arises from intersecting the
LF-affiliated wedge algebras. The full net is only reconstructible if the
action of those remaining Poincar\'{e} transformations outside the
7-parametric LF covariance group is known.

The presence of the Moebius group acting on the lightlike direction on
null-surfaces in curved spacetime resulting from bifurcate Killing horizons
\cite{K-W} has been established in \cite{G-L-W}, thus paving the way for the
transfer of the thermal results to QFT in CST. This is an illustration of
symmetry enhancement which is one of holographies "magics".

The above interaction-free case with its chiral abelian current algebra
structure (\ref{LF}) admits a much larger unitarily implemented symmetry
group, namely the diffeomorphism group of the circle. However the unitary
implementers (beyond the Moebius group) do not leave the vacuum invariant (and
hence are not Wigner symmetries). As a result of the commutation relations
(\ref{com}) these Diff(S$^{1}$) symmetries are expected to appear in the
holographic projection of interacting theories. These unitary symmetries act
only geometrically on the holographic objects; their action on the bulk (on
which they are also well-defined) is fuzzy i.e. not describable in geometric
terms. This looks like an interesting extension of the new setting of local
covariance \cite{B-F-V}

The area proportionality for localization entropy is a structural property of
LQP which creates an interesting and hopefully fruitful contrast with
Bekenstein's are law \cite{Be} for black hole horizons. Bekenstein's thermal
reading of the area behavior of a certain quantity in classical
Einstein-Hilbert like field theories has been interpreted as being on the
interface of QFT with QG. Now we see that the main support, namely the claim
that QFT alone cannot explain an area behavior, is not correct. There remains
the question whether Bekenstein's numerical value, which people tried to
understand in terms of quantum mechanical level occupation, is a credible
candidate for quantum entropy. QFT gives a family of area laws with different
vacuum polarization \textit{attenuation parameters} $\varepsilon$ and it is
easy to fix this parameter in terms of the Planck length so that the two
values coalesce. The problem which I have with such an argument is that I have
never seen a situation where a \textit{classical} value remained intact after
passing to the quantum theory. This does only happen for certain
\textit{quasiclassical} values in case the system is integrable.

\section{From holography to correspondence: the AdS-CFT correspondence and a
controversy}

The holography onto null-surfaces addresses the very subtle relation between
bulk quantum matter and the projection onto its causal/event horizon as
explained in the previous section. A simpler case of holography arises if the
bulk and a lower dimensional brane\footnote{In general the brane has a lower
dimensional symmetry than its associated bulk and usually denotes d-1
dimensional subspace which contains a time-like direction. Different from
null-surfaces branes have a causal leakage.} (timelike) boundary share the
same maximally possible spacetime (vacuum) symmetry. The only case where this
situation arises between two global Lorentz manifolds of different spacetime
dimension is the famous AdS-CFT correspondence. In that case the causality
leakage off a brane does not occur. In the following we will use the same
terminology for the universal coverings of AdS/CFT as for the spacetimes themselves.

Already in the 60s the observation that the 15-parametric conformal symmetry
which is shared between the conformal of 3+1-dimensional compactified
Minkowski spacetime and the 5-dim. Anti-de-Sitter (the negative constant
curvature brother of the cosmologically important de Sitter spacetime) brought
a possible field theoretic relation between these theories into the
foreground; in fact Fronsdal \cite{Fron} suspected that QFTs on both
spacetimes share more than the spacetime symmetry groups. But the modular
localization theory which could convert the shared group symmetry into a
relation between two \textit{different spacetime ordering devices} (in the
sense of Leibniz) for the \textit{same abstract quantum matter substrate} was
not yet in place at that time. Over several decades the main use of the AdS
solution has been (similar to Goedel's cosmological model) to show that
Einstein-Hilbert field equations besides the many desired solution (as the
Robertson-Walker cosmological models and the closely related de Sitter
spacetime) also admit unphysical solutions (leading to timelike selfclosing
worldlines, time machines, wormholes etc.) and therefore should be further restricted.

The AdS spacetime lost this role of only providing counterexamples and began
to play an important role in particle physics when the string theorist placed
it into the center of a conjecture about a correspondence between a particular
maximally supersymmetric massless conformally covariant Yang-Mills model in
d=1+3 and a supersymmetric gravitational model. The first paper was by J.
Maldacena \cite{Ma} who started from a particular compactification of 10-dim.
superstring theory, with 5 uncompactified coordinates forming the AdS
spacetime. Since the mathematics as well as the conceptual structure of string
theory is poorly understood, the string side was identified with one of the
supersymmetric gravity models which inspite of its being non-renormalizable
admitted a more manageable Lagrangian formulation and was expected to have a
similar particle content. On the side of CFT he placed a maximally
supersymmetric gauge theory of which calculations which verify the vanishing
of the low order beta function already existed\footnote{An historically
interesting case in which the beta function vanishes in every order is the
massive Thirring model. In that case the zero mass limit is indeed conformally
invariant, but there is no interacting conformal theory for which a
perturbation can be formulated directly, it would generate unmanagable
infrared divergencies.} (certainly a \textit{necessary} prerequisite for
conformal invariance).\ The arguments involved perturbation theory and
additional less controllable approximations. The more than 4.700 follow up
papers on this subject did essentially not change the status of the
conjecture. But at least some aspects of the general AdS-CFT correspondence
became clearer after Witten \cite{Witten} exemplified the ideas in the field
theoretic context of a $\Phi^{4}$ coupling on AdS using a Euclidean functional
integral setting.

The structural properties of the AdS-CFT correspondence came out clearly in
Rehren's \cite{Rehren} \textit{algebraic holography}. $\ $The setting of local
quantum physics (LQP) is particularly suited for questions in which one theory
is assumed as given and one wants to construct its holographic projection or
its corresponding model on another spacetime. LQP can solve such problems of
isomorphisms between models without being forced to actually construct a model
on either side (which functional integration proposes to do but only in a
metaphoric way) be. At first sight Rehren's setting rewritten in terms of
functional integrals (with all the metaphoric caveats, but done in the best
tradition of the functional trade) looked quite different from Witten's
functional representation. But thanks to a functional identity (explained in
the Duetsch-Rehren paper) which shows that fixing functional sources on a
boundary and forcing the field values to take on a boundary value via delta
function in the functional field space leads to the same result. In this way
the apparent disparity disappeared \cite{Du-Re} and there is only one AdS-CFT
correspondence within QFT.

There are limits to the rigor and validity of functional integral tools in
QFT. Even in QM where they are rigorous an attempt to teach a course on QM
based on functional integrals would end without having been able to cover the
standard material. As an interesting mental exercise just image a scenario
with Feynman before Heisenberg. Since path integral representations are much
closer to the old quasiclassical Bohr Sommerfeld formulation the transition
would have been much smoother, but it would have taken a longer time to get to
the operational core of quantum theory; on the other hand quasiclassical
fomulas and perturbative corrections thereof would emerge with elegance and
efficiency. 

Using the measure theoretical functional setting it is well-known that
superrenormalizable polynomial couplings can be controlled this way
\cite{Gl-Ja}. Realistic models with infinite wave function renormalization
constants (all realistic Lagrangian models in more than two spacetime
dimensions have a trans canonical short distance behavior) do not fall into
this amenable category. But even in low dimension, where there exist models
with finite wave function renormalization constants and hence the short
distance prerequisites are met, the functional setting of the AdS-CFT
correspondence has an infrared problem\footnote{Infrared problems of the kind
as they appear in interacting conformal theories are strictly speaking not
susceptible to perturbation theoretical treatment and they also seem to pose
serious (maybe unsoluble) problems in functional integral representations. In
those cases where on knows the exact form of the massless limit (Thirring
model) this knowledge can be used to disentangle the perturbative infrared
divergences.} of a nasty unresolved kind \cite{Got}. As the result of lack of
an analog to the operator formulation in QM the suggestive power, their close
relation to classical geometric concepts and their formal elegance functional
integrals have maintained their dominant role in particle physics although
renormalized perturbation theory is better taken care of in the setting of
"causal perturbation". An operator approach which is not only capable to
establish the mathematical existence of models but also permits their explicit
construction exists presently only in d=1+1; it is the previously mentioned
bootstrap-formfactor  or wedge-localization approach for fsctorizing models.
Lagrangian factorizing models only constitute a small fraction. 

For structural problems as holography, where one starts from a given theory
and wants to construct its intrinsically defined holographic image, the use of
metaphorical instruments as Euclidean functional integral representations is
suggestive but not really convincing in any mathematical sense. As in the case
of lightfront holography there are two mathematically controllable ways to
AdS-CFT holography; either using (Wightman) fields (\textit{projective
holography}) or using operator algebras (\textit{algebraic holography}). The
result of all these different methods can be consistently related
\cite{Du-Re}\cite{ReLec}.

The main gain in lightfront holography is a significant simplification of
certain properties as compared to the bulk. Even if some of the original
problems of the bulk come back in the process of holographic inversion they
reappear in the more amenable form of several smaller problems rather than one
big one. 

The motivation for field theorists being interested in the AdS-CFT
correspondence is similar, apart from the fact that the simplification
obtainable through an \textit{algebraic isomorphism} is more limited (less
radical) than that of a projection. Nevertheless it is not unreasonable to
explore the possibility whether some hidden property as e.g. a widespread
conjectures \textit{partial integrabilty}\footnote{Global integrability is
only possible in d=1+1, but I am not aware of any theorem which rules out the
possibility of integrable substructures. } could become more visible after a
spacetime "re-packaging" of the quantum matter substrate from CFT to AdS. 

Despite many interesting analogies between chiral theories and higher
dimensional QFT \cite{To} little is known about higher-dimensional conformal
QFTs. There are Lagrangian candidates as e.g. certain supersymmetric
Yang-Mills theories which fulfill (at least in lowest order) some perturbative
prerequisite of conformality which consists in a vanishing beta-function. As
mentioned before perturbation theory for conformal QFT, as a result of severe
infrared problems,  cannot be formulated directly. The prime example for such
a situation is the massive Thirring model for which there exists an elegant
structural argument for $\beta(g)=0$ and the knowledge about the
non-perturbative massless version can then be used to find the correct
perturbative infrared treatment. 

As far as I could see (with appologies in case of having overlooked some
important work) none of these two steps has been carried out for SUSY-YM, so
even the conformal side of the Maldacena conjecture has remained unsafe
territory. 

There is one advantage which null-surface holography has over AdS-CFT type
brane holography. The cardinality of degrees of freedom adjusts itself to what
is \textit{natural} for null-surfaces (as a manifold in its own right); for
the lightfront holography this is the operator algebra generated from extended
chiral fields (\ref{com}). On the other hand this "thinning out" in
holographic projections is of course the reason whay inverse holography
becomes more complicated and cannot be done with the QFT on one null surface only.

In the holography of the AdS-CFT correspondence the bulk degrees of freedom
pass to a conformal brane; in contradistinction to the holography on
null-surfaces there is \textit{no reduction of degrees of freedom} resulting
from projection. \ Hence the AdS$-$CFT isomorphism starting from a "normal"
(causally complete as formally arising from Lagrangians) 5-dimensional AdS
leads to a \textit{conformal field theory with too many degrees of freedom}.
Since a "thinning out" by hand does not seem to be possible, the "physically
health" of such a conformal QFT is somewhat dodgy, to put it mildly. 

In case one starts with a free Klein-Gordon field on AdS one finds that the
generating conformal fields of the CFT are special \textit{generalized free
fields} i.e. a kind of continuous superpositions of free fields. They were
introduced in the late 50s by W. Greenberg and their useful purpose was
(similar to AdS in classical gravity) to \textit{test the physical soundness
of axioms of QFT} in the sense that if a system of axioms allowed such
solutions, it needed to be further restricted \cite{Ha-Sc} (in that case the
so-called causal completion or time-slice property excluded generalized free
fields). It seems that meanwhile the word "physical" has changes its meaning,
it is used for anything which originated from a physicist.

In the opposite direction the degrees of freedom of a "normal" CFT become
"diluted" on AdS in the inverse correspondence. There are not sufficient
degrees of freedom for arriving at nontrivial compactly localized operators,
the cardinality of degrees of freedom is only sufficient to furnish noncompact
regions as AdS wedges with nontrivial operators, the compactly localized
double cone algebras remain trivial (multiples of the identity). In the
setting based on fields this means that the restriction on testfunction spaces
is so severe that pointlike field $A_{AdS}(x)$ at interior points $x\in
intAdS$ do not exist in the standard sense as operator-valued distributions on
Schwartz spaces. They exist on much smaller test function spaces which contain
no functions with compact localizations.

Both sides of the correspondence have been treated in a mathematically
rigorous fashion for free AdS (Klein-Gordon equation) theories and free (wave
equation) CFT \cite{Du-Re2}\cite{ReLec} where the mismatch between degrees of
freedom can be explicated and the structural arguments based on the principles
of general QFT show that this mismatch between the transferred and the natural
cardinality of the degree of freedom is really there. In terms of the better
known Lagrangian formalism the statement would be that if one starts from a
Lagrange theory at one side the other side cannot be Lagrangian. Of course
both sides remain QFT in the more general sense of fulfilling the required
symmetries, have positive energy and being consistent with spacelike
commutativity. In the mentioned free field illustration a AdS Klein-Gordon
field is evidently Lagrangian whereas the corresponding \textit{conformal
generalized free field} has no Lagrangian and cannot even be characterized in
terms of a local hyperbolic field equation. According to the best educated
guess,  4-dim. maximally supersymmetric Yang-Mills theories (if they exist and
are conformal) would be a natural conformal QFTs "as we know it" and therefore
cannot come from a natural QFT on AdS. Needless to say again that there are
severe technical problems to set up a perturbation theory for a conformally
invariant interactions, the known perturbative systematics breaks down in the
presence of infrared problems\footnote{A well-known problem is the massive
Thirring model which leads to $\beta=0$ in all orders. In this case one
already knew confmal limit in closed form and was able to check the
correctness of the relation by consistency considerations.}.

I belong to a generation for which not everything which is mathematically
possible must have a physical realization; in particular I do not adhere to
the new credo that every mathematically consistent idea is realized in some
parallel world (anthropic principle): no parallel universe for the physical
realization of every mathematical belch. 

Generalized free fields\footnote{It is interesting to note that the Nambu-Goto
Lagrangian (which describes a classical relativistic string) yields upon
quantization a pointlike localized generalized free field with the well-known
infinite tower mass spectrum and the appearance of a Hagedorn limit
temperature. As such it is pointlike localized and there is \textit{no
intrinsic quantum concept} which permits to associate it with any stringlike
localization. } and their interacting counterparts which arise from natural
AdS free- or interacting- fields remain in my view unphysical, but are of
considerable mathematical interest. They do not fit into the standard causal
localization setting and they do not allow thermal KMS states without a
limiting Hagedorn temperature (both facts are related). Nature did not
indicate that it likes to go beyond the usual localizability and thermal
behavior. If string theory demands such things it is not my concern, let Max
Tegmark find another universe where nature complies with string theory. \ 

Holography is a technical tool and not a physical principle. It simplifies
certain aspects of a QFT at the expense of others (i.e. it cannot achieve
miracles). The use of such ideas in intermediate steps may have some technical
merits, but I do not see any scientific reason to change my viewpoint about
physical admissibility. The question of whether by changing the spacetime
encoding one could simplify certain properties (e.g. detect integrable
substructures) of complicated theories is of course very interesting, but in
order to pursue such a line it is not necessary to physically identify the
changed theory. Such attempts where only one side needs to be physical and the
role of holography would consist in exposing certain structural features which
remained hidden in the original formulation sound highly interesting to me.

There is however one deeply worrisome aspect of this whole development. Never
before has there been more than 4.700 publication on such a rather narrow
subject; in fact even nowadays, one decade after this gold-digger's rush about
the AdS-CFT correspondence started, there is still a sizable number of papers
every month by people looking for nuggets at the same place but without
bringing Maldacena's gravity-gauge theory conjecture any closer to a
resolution. Even with making all the allowances in comparison with earlier
fashions, this phenomenon is too overwhelming order to be overlooked.
Independent of its significance for particle physics and the way it will end,
the understanding of what went on and its covering by the media will be
challenging to historians and philosophers of science in the years to come. 

I know that it is contra bonos mores to touch on a sociological aspect in a
physics paper, but my age permits me to say that at no time before was the
scientific production in particle theory that strongly coupled to the
Zeitgeist as during the last two decades; never before had global market
forces such a decisive impact on the scientific production. Therefore it is
natural to look for an explanation why thousands of articles are written on an
interesting (but not clearly formulated) conjecture with hundreds of other
interesting problems left aside; where does the magic attraction come from? Is
it the Holy Grail of a TOE which sets into motion these big caravans? Did the
critical power of past particle physics disappear in favor of acclamation? Why
are the few critical but unbiased attempts only mentioned by the labels given
to them and not by their scientific content?

Since commentaries about the crisis in an area of which one is part run the
risk of being misunderstood, let me make perfectly clear that particle physics
was a speculative subject and I uphold that it must remain this way. Therefore
I have no problem whatsoever with Maldacena's paper; it is in the best
tradition of particle physics which was always a delicate blend of a highly
imaginative and innovative contribution from one author with profoundly
critical analysis of others. I am worried about the loss of this balance. My
criticism is also not directed against the thousands of authors who enter this
area in good faith believing that they are working at an epoch-forming
paradigmatic problem because their peers gave them this impression. Even if
they entered for the more mundane reason of carving out a career, I would not
consider this as the cause of the present problem. 

The real problem is with those who by their scientific qualifications and
status are the intellectual leaders and the role models. If they abdicate
their role as critical mediators by becoming the whips of the TOE monoculture
of particle physics then checks and balances will be lost. Would there have
been almost 5000 publication on a rather narrow theme (compared with other
topics) in the presence of a more critical attitude from leading particle
physicists? No way. Would particle theory, once the pride of theoretical
physics with a methodological impact on many adjacent areas have fallen into
disrespect and be the object of mock within the larger physics community? The
list of questions of this kind with negative answers can be continued.

It is worthwile to look back at times when the delicate balance between the
innovative and speculative on the one hand and the critical on the other was
still there. Young researchers found guidance by associating themselves to
\ "schools of thought" which where associated with geographical places and
names as Schwinger, Landau, Bogoiubov, Wheeler, Wightman, Lehmann, Haag... who
represented different coexisting schools of thought. Instead of scientific
cross fertilization between different schools, the new globalized caravan
supports the formation of a gigantic monoculture and the loss of the culture
of checks and balances.

Not even string theorists can deny that this unfortunate development started
with string theory. Every problem string theory addresses takes on a strange
metaphoric aspect, an effect which is obviously wanted as the fondness for the
use of the letter M shows. The above mentioned AdS-CFT topic gives an
illustration which (with a modest amount of mathematical physics) shows the
clear structural QFT theorem as compared to the strange conjecture which even
thousands of publications were not able to liberate from the metaphoric
twilight. 

But it is a remarkable fact that, whenever string theorist explain their ideas
by QFT analogs in the setting of functional integrals as was done by Witten in
\cite{Witten} for the $\varphi^{4}$ coupling, and on the other hand algebraic
quantum field theorists present their rigorous structural method for the same
model in the same setting \cite{Du-Re}, the two results agree (see also
\cite{Got}). 

This is good news. But now comes the bad news. Despite the agreement the
Witten camp, i.e. everybody except a few individuals, claim that there exist
two different types of AdS-CFT correspondences namely theirs and another one
which at least some of them refer to as the "German AdS-CFT correspondence".
Why is that? I think I know but I will not write it.

At this point it becomes clear that it is the abandonment of the critical role
of the leaders which is fuelling this unhealthy development. Could a
statement: \ "X-Y-Z theory is a gift of the 21st century which by chance fell
into the 20 century" have come from Pauli, Schwinger, or Feynman? One would
imagine that in those days people had a better awareness that mystifications
like this could disturb the delicate critical counterbalance which the
speculative nature of particle physics requires. \ The long range negative
effect on particle theory of such a statement is proportional to the
prominence and charisma of its author. 

There have been several books which criticise string theory. Most critics
emphasize that the theory has not predicted a single observable effect and
that there is no reason to expect that this will change in the future.
Although I sympathize with that criticism, especially if it comes from
experimentalists and philosphers, I think that a theorist should focus his
critique on the conceptual and mathematical structure and not rely on help
from Karl Popper or dwell on the non-existent observational support.
Surprisingly I could not find any scholarly article in this direction. One of
the reasons may be that after 4 decades of development of string theory such a
task requires rather detailed knowledge about its conceptual and mathematical
basis. As a result of this unsatidfactory situation I stopped my critical
article \cite{crisis} from going into print and decided to re-write it in such
a way that the particle physics part is strengthened at the expense of the
sociological sections. 

The aforementioned situation of ignoring results which shed a critical light
on string theory or the string theorists version of the AdS-CFT correspondence
is perhaps best understood in terms of the proverbial \textit{executing of the
messenger who brings bad news}; the unwanted message in the case at hand being
the \textit{structural} impossibility to have Lagrangian QFTs with causal
propagation on both sides of the correspondence.

It seems that under the corrosive influence of more than 4 decades of string
theory, Feynman's observation about its mode of arguing being based on finding
excuses instead of explanations, which two decades ago was meant to be
provocative, has become the norm. The quantum gravity-gauge theory conjecture
is a good example of how a correct but undesired AdS-CFT correspondence is
shifted to the elusive level of string theory and quantum gravity so that the
degrees of freedom aspect becomes pushed underneath the rug of the elusive
string theory where it only insignificantly enlarges the already very high
number of metaphors. 

There have been an increasing number of papers with titles as "QCD and a
Holographic Model of Hadrons", "Early Time Dynamics in Heavy Ion Collisions
and AdS/CFT Correspondence", "Confinement/Deconfinement Transition in AdS /
CFT", "Isospin Diffusion in Thermal AdS/CFT with flavour", "Holographic Mesons
in a Thermal Bath", "Viscous Hyrodynamics and AdS/CFT", "Heavy Quark Diffusion
from AdS/CFT".... Ads/CFT \ for everything? Is string theory bolstered by
AdS-CFT really on the way to become a TOE for all of physics, a theory for
anything which sacrifies conceptual cohesion to amok running calculations? Or
are we witnessing a desperate attempt to overcome the more than 4 decade
lasting physical disutiliy? Perhaps it is only a consequence of the
"liberating" effect of following prominent leaders who have forgone their duty
as critical mediators and preserver of conceptual cohesion. 

\section{Concluding remarks}

In these notes we revisited one of the oldest and still unsolved conceptual
problems in QFT, the existence of interacting models. Besides some new
concrete results about the existence of factorizing models (which only exist
in d=1+1), it is the new method itself, with its promise to explore new
fundamental and fully intrinsic properties of QFT, which merits attention. A
particularly promising approach for the classification and construction of
QFTs consists in using holographic lightfront projections (and in a later
stage work one's way back into the bulk). In this situation the holographic
degrees of freedom are thinned out as compared to the bulk i.e. the extended
chiral fields have lesser number of degrees of freedom.

The concept of degrees of freedom used here is a dynamical one. Knowing only a
global algebra\footnote{Knowing an operator algebra means knowing its position
within the algebra $B(H)$ of all operators. Knowing its net substructure means
knowing the relative position of all its subalgebras.} as the wedge algebra
i.e. $\mathcal{A}(W)\subset B(H)$ as an inclusion into the full algebra one
uses fewer degrees freedom than one needs in order to describe the full local
substructure of $\mathcal{A}(W)$ i.e. knowing $\mathcal{A}(W\not )$ in the
sense of a local net. The degrees of freedom emerge always from relations
between algebras whereas the single algebra is a structureless monad\cite{Hol}%
. \ Saying that the net $\mathcal{A}(LF)$ has less degrees of freedom than the
net associated with the bulk is the same as saying that the knowledge of the
$LF$affiliated wedges does not suffice to reconstruct the local bulk
structure. In this sense the notion of degrees of freedom depends on the
knowlege one has about a system; refining the net structure of localized
subalgebras of a global algebra increases the degrees of freedom.  

The lightfront holography is a genuine projection with a lesser cardinality of
degrees of freedom i.e. without knowing how other Poincar\'{e} transformations
outside the 7-parametric invariance group of the lightfront act it is not
uniquely invertible. On its own, i.e. without added information, the
lightfront holography cannot distinguish between massive and massless
theories; \ a transverse extended chiral theories does not know whether the
bulk was massive or massless. The knowledge of how the opposite lightray
translation $U(a_{-})$ acts on $\mathcal{A}(LF)$ restores uniqueness; but this
action is necessarily "fuzzy" i.e. non-geometric, purely algebraic$.$Only upon
returning to the spacetime ordering device in terms of the bulk it becomes geometric.

The hallmark of null-surface holography is an area law for localization
entropy in which the proportionality constant is a product of a holographic
matter dependent constant times a logarithmic dependence on the attenuation
length for vacuum polarization.

By far the more popular holography has been the AdS-CFT correspondence. Here
its physical utility is less clear than the mathematical structure. 

Is there really a relation between a special class of conformal gauge
invariant gauge theories with supersymmetric quantum gravity? Not a very
probable consequence of a change of an spacetime ordering device for a given
matter substrate which is what holography means. Integrable substructures
within such conformal gauge theories which become more overt on the AdSside?
This appears a bit more realistic, but present indications are still very flimsy.

\textbf{Acknowledgements}: I am indebted to B. Fauser, J. Tolksdorf and E.
Zeidler for the invitation to participate in a 2007 conference in Leipzig and
for hospitality extended to me during my stay.

.
\end{document}